\documentclass{article}
\usepackage{spconf,amsmath,graphicx,hyperref}

\title{Reference-Aware SFM Layers for Intrusive Intelligibility Prediction}
\name{%
  \begin{tabular}{c}
    Hanlin Yu$^{1}$, Haoshuai Zhou$^{2}$, Boxuan Cao$^{2}$,\\
    Changgeng Mo$^{2}$, Linkai Li$^{2,3}$, Shan X.~Wang$^{3}$\thanks{Correspondence: \texttt{linkaili@stanford.edu}, \texttt{sxwang@stanford.edu}.}
  \end{tabular}
}

\address{%
$^{1}$ UBC ECE, Canada \quad
$^{2}$ Orka Labs Inc., China \quad
$^{3}$ Stanford EE, USA
}

\begin{document}
%\ninept
%
\maketitle
\noindent\footnotesize\textit{Preprint—submitted to ICASSP 2026.}\par\normalsize
\begin{abstract}
Intrusive speech-intelligibility predictors that exploit explicit reference signals are now widespread, yet they have not consistently surpassed non-intrusive systems. We argue that a primary cause is the limited exploitation of speech foundation models (SFMs). This work revisits intrusive prediction by combining reference conditioning with multi-layer SFM representations. 
 Our final system achieves RMSE 22.36 on the development set and 24.98 on the evaluation set, ranking 1st on CPC3. These findings provide practical guidance for constructing SFM-based intrusive intelligibility predictors.
\end{abstract}
\begin{keywords}
speech intelligibility prediction, intrusive modeling, hearing-impaired listeners
\end{keywords}

\section{Introduction}
Accurate prediction of speech intelligibility is valuable for selecting signal–processing strategies, tuning hearing-aid parameters, and benchmarking enhancement systems for hearing-impaired listeners. Classical intrusive metrics compare a processed signal to a clean reference and quantify envelope or spectro–temporal similarity, whereas non-intrusive predictors estimate intelligibility directly from the degraded waveform or features~\cite{taal2011stoi}. Recent progress has been driven by speech foundation models (SFMs~\cite{cuervo2024sfm}), whose internal representations capture rich phonetic and lexical regularities and have proven effective for intelligibility prediction. Intelligibility hinges on whether a listener can recover the linguistic content of a signal. ASR encoders are trained precisely for this acoustics-to-text mapping, so their hidden states form a natural feature space for intelligibility prediction. We therefore adopt SOTA ASR encoders—Canary-1B-flash~\cite{canary1bflash_modelcard} and parakeet-tdt-0.6b-v2~\cite{parakeet06bv2_modelcard} as our SFM backbone and study reference-conditioned, multi-layer aggregation.

We integrate explicit reference information within an SFM-based pipeline to improve performance. We attribute this, in part, to the under-utilization of speech foundation models (SFMs) within intrusive pipelines. Instead of operating on raw waveforms or shallow hand-crafted features (e.g., STOI with phonetic cues\cite{cpc2_e009_huckvale2023}, ASR hidden/decoder similarity\cite{cpc2_e022_tu2023}, or STM-based CNNs\cite{cpc3_e020a_zhou2025}), we feed the clean reference to an SFM and perform reference-conditioned, multi-layer aggregation to expose phonetic, lexical, and prosodic priors directly relevant to intelligibility.

In this work we systematically examine key design choices for SFM-based intrusive prediction: (i) reference-conditioned versus reference-free training; (ii) “best ear” versus binaural averaging; (iii) which SFM layers to aggregate and how; (iv) conditioning on severity versus continuous audiograms.

\section{Model}
\begin{figure*}[!t]
  \centering
  \includegraphics[width=.90\textwidth]{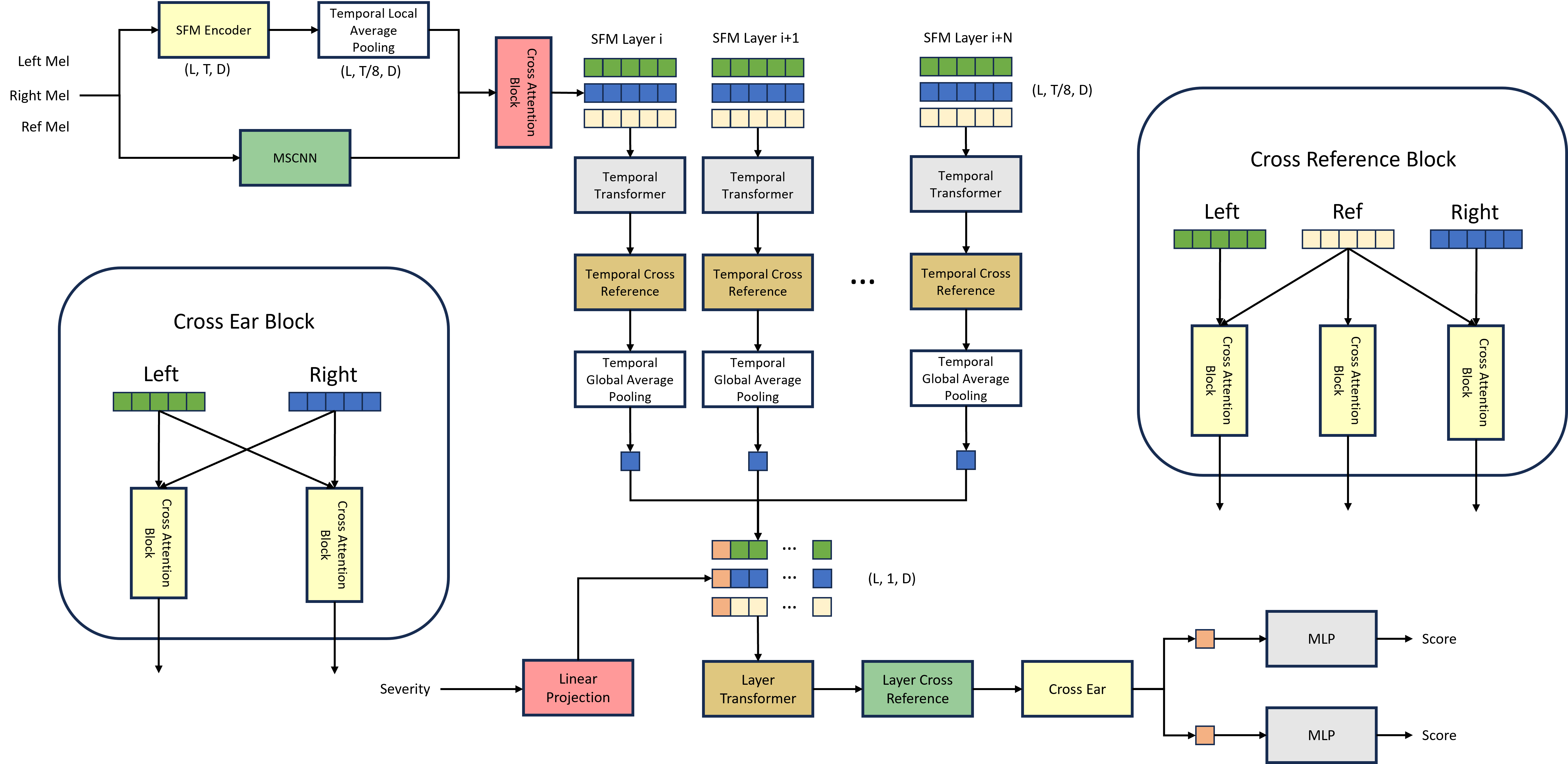}  
\caption{\textbf{Model overview.} Left, right, and reference streams use SFM features fused with an MSCNN front end, then pass through a depth-1 temporal transformer and a depth-1 layer transformer (weights shared across streams). Cross-reference attention is applied after the temporal and layer transformers, and cross-ear attention after the layer transformer. Finally, left/right ear scores are produced by a shared MLP and combined with a temperature-controlled log-sum-exp (softmax) pooling, yielding a differentiable “best-ear” utterance score ($\beta{=}6$).}

  \label{fig:overview}
  \vspace{-2mm}
\end{figure*}

Figure \ref{fig:overview} shows our pipeline: L/R HA signals and the clean reference are encoded by SFM layers, then each ear attends to the reference and to the other ear. The severity tokens go through a shared MLP head and softmax (log-sum-exp) best-ear pooling to yield the utterance-level score.

\subsection{Problem Setup}
We work under the Clarity Prediction Challenge 3 (CPC3) protocol~\cite{cpc3_intro}: predict utterance-level intelligibility for binaural, hearing-aid processed speech in noise, with systems ranked by RMSE on the evaluation split~\cite{cpc3_results}. We adopt the official Train/Dev splits for development and report final numbers on Eval. Split sizes are: Train 15{,}464 signals from 18 enhancement systems and 26 listeners (WHO severities 9/13/4 for Mild/Moderate/Moderately severe), Dev 926 from 4 systems and 8 listeners (3/4/1), and Eval 7{,}674 from 9 systems and 16 listeners (5/10/1).

Our model takes the left/right HA-processed signals and a clean reference, together with a listener severity label (mild, moderate, or moderately severe). It outputs a score in $[0,100]$ and is trained to minimize RMSE to the ground-truth intelligibility.

\subsection{Model Structure: SFM–MSCNN Fusion}
We keep the mid–deep encoder layers (10–16) from both backbones (Canary-1B-flash and parakeet-tdt-0.6b-v2), following Sec.~\ref{Find best layers}.
For each stream (Left/Right/Reference) , we extract the selected SFM layers and apply $\times 8$ temporal average pooling.

From log-Mel features $[B,T,128]$, a three-branch dilated 1-D CNN (kernels $3/5/9$, dilations $1/2/4$, SAME padding, stride $=1$) produces frame-level embeddings
without downsampling. These full-rate MSCNN features serve as keys/values in a cross-attention block (see Sec.~\ref{sec:crossattn}) where downsampled SFM tokens are the queries
implemented with our cross attention block.
Weights are shared across streams and layers. The fused outputs (one per selected layer) form the per-stream layer sequence passed to the subsequent temporal/layer transformers. This step injects fine-grained acoustics into SFM tokens, boosting robustness.

\subsection{Cross-Attention Blocks and Cross-Stream Fusion}
\label{sec:crossattn}
\noindent\textbf{CrossAttentionBlock.}

Given queries $Q$, keys $K$, values $V$ and an optional mask $M$, we apply multi-head attention,
then residual + LayerNorm, followed by a gated feed-forward.
Here $\mathrm{FFN}$ is a two-layer MLP with SiLU and GLU gates and dropout.
This block is used (i) to fuse downsampled SFM tokens with full-rate MSCNN features (SFM–MSCNN fusion), (ii) to exchange information between each ear and the reference stream, and (iii) to exchange information between the two ears.

\noindent\textbf{Temporal stage (within layer).}
After SFM features and MSCNN embeddings are fused at each selected layer, the tokens are passed through a depth-1 temporal transformer that models dependencies along the time axis. 
This allows each SFM layer to capture temporal context beyond frame-level information. 
At the same stage, the left- and right-ear streams are aligned with the clean reference through cross-attention, so that each ear representation can directly attend to reference cues from the same temporal window. 
A masked mean over time then produces one summary vector per layer and per stream.

\noindent\textbf{Layer stage (across layers).}
The per-layer vectors are then stacked into a sequence that represents the progression of features from lower to higher SFM layers. 
A depth-1 layer transformer models the inter-layer relationships, i.e., how mid-deep layers complement each other in representing phonetic and lexical information. 
At this stage, each ear again attends to the reference sequence (ear–reference cross-attention) so that higher-level priors from the reference are integrated. 
Finally, cross-ear attention exchanges information between the two ears, enabling the model to weight the better ear more heavily while still leveraging binaural cues. 

\section{Experiments}

\subsection{Feature Extractors and Layer Aggregation}
\label{Find best layers}
\noindent\textbf{Backbone}.
We use Canary-1B-flash as the SFM encoder and expose all 32 hidden layers. Encoders are frozen for these sweeps, and all other hyperparameters are kept constant. We fix the listener split on train dataset: 6 listeners form the validation set; the rest are used for training.

\noindent\textbf{Motivation and setups.}
Recent CPC-style work shows two effective ways to use SFM layers for intelligibility prediction:
(i) using a CLS token so that the model can learn weighted summaries across time/layers \cite{cuervo2024sfm};
(ii) selecting specific encoder layers rather than all layers, which often yields better accuracy \cite{zhou2025bestpractices}.
Guided by these findings, we compare three variants under our intrusive architecture (All three variants share the same backbone as Fig.~\ref{fig:overview} (binaural SFM encoder, reference conditioning, temporal and layer transformers, cross-ref and cross-ear); the only difference lies in how the ear representations are read out):

\emph{Setup A (ours)}: Append one severity token to each layer-token sequence for left, right, and reference branches. These sequences go through the layer-wise Transformer then cross-ref then cross-ear. After those fusions, take the severity token position from each ear branch and feed them to the shared MLP for scores.

\emph{Setup B (mean pooling)}:
We append a severity token to each per-layer token sequence for the left, right, and reference streams and process them with layer Transformer, followed by the same cross-reference and cross-ear modules as in Setup A. For scoring, take a mean across the selected layer window to obtain a single vector, which is fed to the shared MLP (mean-pooling baseline following\cite{zhou2025bestpractices}).

\emph{Setup C (CLS pooling)}:
We insert a learnable \texttt{[CLS]} token at both stages. At the \textit{temporal} stage, each layer sequence includes a CLS whose output after the temporal transformer (and cross-reference) serves as that layer’s summary. These per-layer summaries are then stacked; a second CLS is prepended at the \textit{layer} stage, and the layer Transformer outputs a layer-CLS vector. This layer-CLS is taken as the ear representation (no mean pooling) and passed to the shared MLP (CPC2-style CLS pooling following\cite{cuervo2024sfm}). When a severity token is present, it participates via attention but is not used as the readout vector.

\noindent\textbf{How we select “best layer(s)”.}
We sweep either (i) a single layer (\emph{window size} = 1) or (ii) contiguous blocks of 4 layers (win=4). For each candidate we train and record best validation RMSE; the lowest RMSE determines the chosen layer/block.

\begin{figure}[t!]
  \centering
  \includegraphics[width=.70\columnwidth]{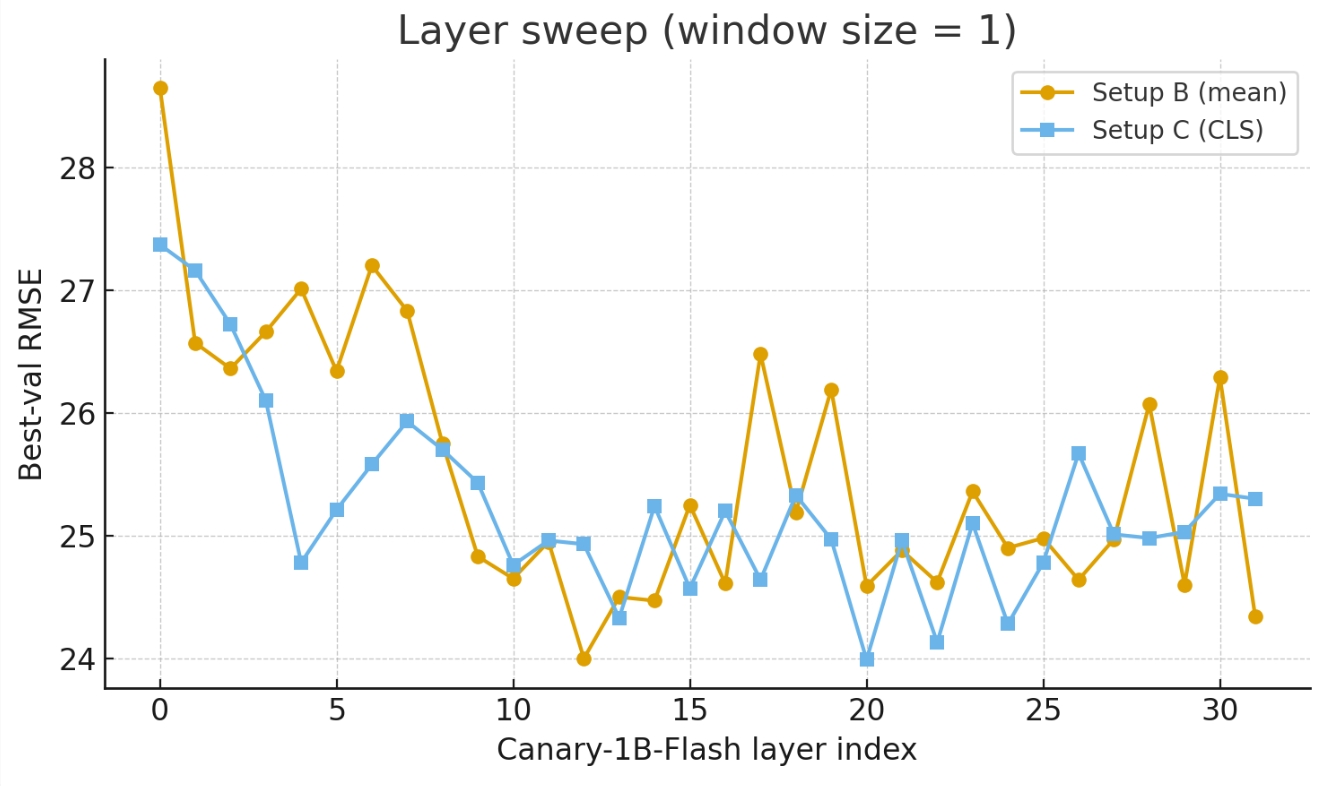}
  \caption{Canary-1B-flash layer sweep on Canary-1B-flash (window size = 1). Setup C (CLS) and Setup B (mean); best at L20 (23.99) vs. L12 (24.00).}
  \label{fig:layer-sweep-ws1}
  \vspace{-2mm}
\end{figure}

\noindent\textbf{Single-layer sweep (win=1).}
Fig.~\ref{fig:layer-sweep-ws1} shows per-layer RMSE. We omit Setup A (severity-token readout) for win=1: with only one SFM layer, the severity token becomes the sole readout and can overpower audio evidence, biasing the readout comparison.
Setup C (CLS) outperforms Setup B (mean), with best layers at L20 (23.99) vs. L12 (24.00), indicating CLS summarizes information more effectively than uniform averaging.

\begin{figure}[t!]
  \centering
  \includegraphics[width=.70\columnwidth]{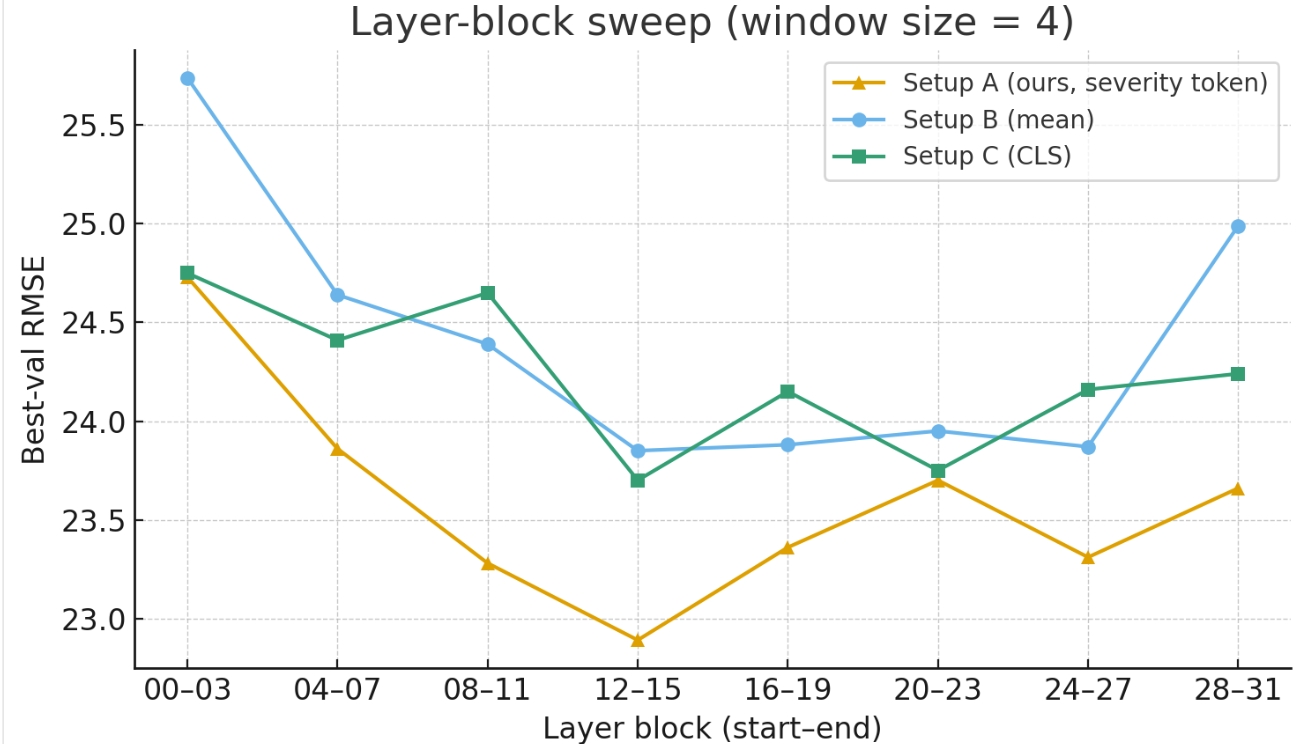}
  \caption{Canary-1B-flash layer-block sweep (window size = 4). Best block is 12–15 for all setups; Setup A achieves 22.89 while Setup B is 23.85 and Setup C is 23.70.}
  \label{fig:layer-sweep-ws4}
  \vspace{-2mm}
\end{figure}

\noindent\textbf{Four-layer blocks (win=4).}
Fig.~\ref{fig:layer-sweep-ws4} compares A/B/C across 4-layer windows. All methods prefer mid-deep layers (12–15). Setup A (ours, severity token) yields the best RMSE (22.89) on 12–15, surpassing Setup C (23.70) and Setup B (23.85). Aggregating more layers helps, and explicit listener conditioning (severity token) gives the strongest gains.

\begin{figure}[t]
  \centering
  \includegraphics[width=.70\columnwidth]{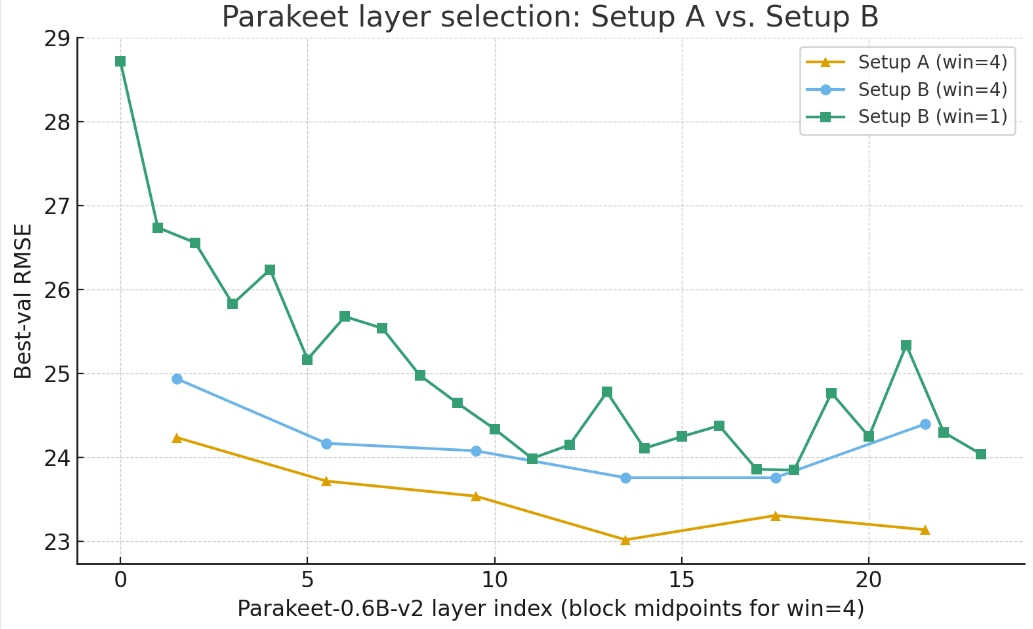}
  \caption{parakeet-tdt-0.6b-v2 layer selection. Setup~A (win=4) and mean-pooling baselines both favor mid–deep layers, peaking at 12–15 (23.02).}
  \label{fig:Parakeetsweep}
  \vspace{-2mm}
\end{figure}

\noindent\textbf{Takeaway.}
Across both SFMs, mid–deep layers are most effective for multi-layer, and listener severity tokens
further improve robustness and accuracy versus mean or CLS pooling alone. 
On Canary this coincides with Setup~A (severity token) outperforming mean and CLS pooling. Because Canary and Parakeet share very similar FastConformer encoder backbones (both NVIDIA SFMs), both encoders expose 1024\,-dim hidden states, which lets us aggregate layers across the two backbones without extra projection, yielding a richer layerwise ensemble.
 We avoided redundant sweeps on Parakeet: we restricted the ablation to Setup~A vs.\ mean pooling. The Parakeet sweep likewise peaks at 12--15 (Fig.~\ref{fig:Parakeetsweep}), and Setup~A is best within that window. We therefore use a slightly wider band—\textbf{layers 10–16} from both Canary and Parakeet—in the final model to capture more information.

\subsection{Setup}
\noindent\textbf{Folds.}
We use 5-fold listener-level CV on the CPC3 train split. In each fold, the validation set contains 6 listeners (2 mild, 2 moderate, 2 moderately severe); the rest are used for training. The checkpoint with the lowest validation RMSE is kept.

\noindent\textbf{Inference.}
For Dev and Eval, we run all five fold-specific checkpoints and average their per-utterance predictions.

\noindent\textbf{Training.}
We train with the AdamW optimizer (learning rate $3\times10^{-5}$, weight decay $10^{-2}$) using a batch size of 8 for 9 epochs with mixed precision. Best-ear temperature $\beta{=}6$. SFMs are frozen; layers 10–16 from both backbones are used.

\subsection{Results and Discussion}

\begin{table}[t]
\centering
\caption{CPC3 results (RMSE; lower is better).}
\label{tab:results}
\begin{tabular}{lcc}
\hline
Method / Variant & Dev & Eval \\
\hline
HASPI baseline & 28.00 & 29.50 \\
\textbf{Ours (CPC3 Champion)} & 22.36 & \textbf{24.98} \\
CPC3 Second place~\cite{cpc3_results} & \textbf{21.87} & 25.31 \\
CPC3 Third place~\cite{cpc3_results}  & 22.80 & 25.54 \\
\hline
Severity: PTA4 (replace token) & \textbf{22.29} & 25.11 \\
Severity: PTA8 (250--8000 Hz) & 23.20 & 25.30 \\
No severity (CLS instead) & 23.88 & 25.69 \\
No reference & 22.82 & 25.39 \\
Average ear (feature fuse) & 22.82 & 25.29 \\
\hline
\end{tabular}
\end{table}

\noindent\textbf{Discussion.}
\textbf{(1) Severity conditioning.} CPC3 severity labels follow the WHO rule based on \emph{PTA4}, i.e., the average audiometric threshold at 0.5, 1, 2, and 4 kHz. We therefore compare three encodings: (i) a categorical severity token; (ii) \emph{PTA4} values projected by a linear layer; (iii) \emph{PTA8} using 250–8000 Hz (8 bands). As Table~\ref{tab:results} shows, PTA4 performs on par with the severity token, but PTA8 degrades—higher bands add noise/irrelevant variation for this task. Removing severity altogether (replacing it with a generic CLS) hurts the most, confirming the importance of listener conditioning.

\textbf{(2) Reference usage.} In the no reference ablation we remove the entire reference stream and disable the cross-reference modules; all other components remain unchanged. RMSE increases on both Dev and Eval, indicating that explicit reference conditioning is beneficial even with strong SFMs.

\textbf{(3) Ear pooling.} Our default uses temperature-controlled log-sum-exp (“best-ear”) to combine left/right scores. The “average-ear” baseline follows the CPC2-winning practice \cite{cuervo2024sfm}: after cross-ear, we average the left/right ear representations at the feature level to obtain a single vector, then apply the shared MLP. Average-ear underperforms best-ear, consistent with the intuition that intelligibility is dominated by the better ear.
\begin{figure}[t]
  \centering
  \includegraphics[width=\columnwidth]{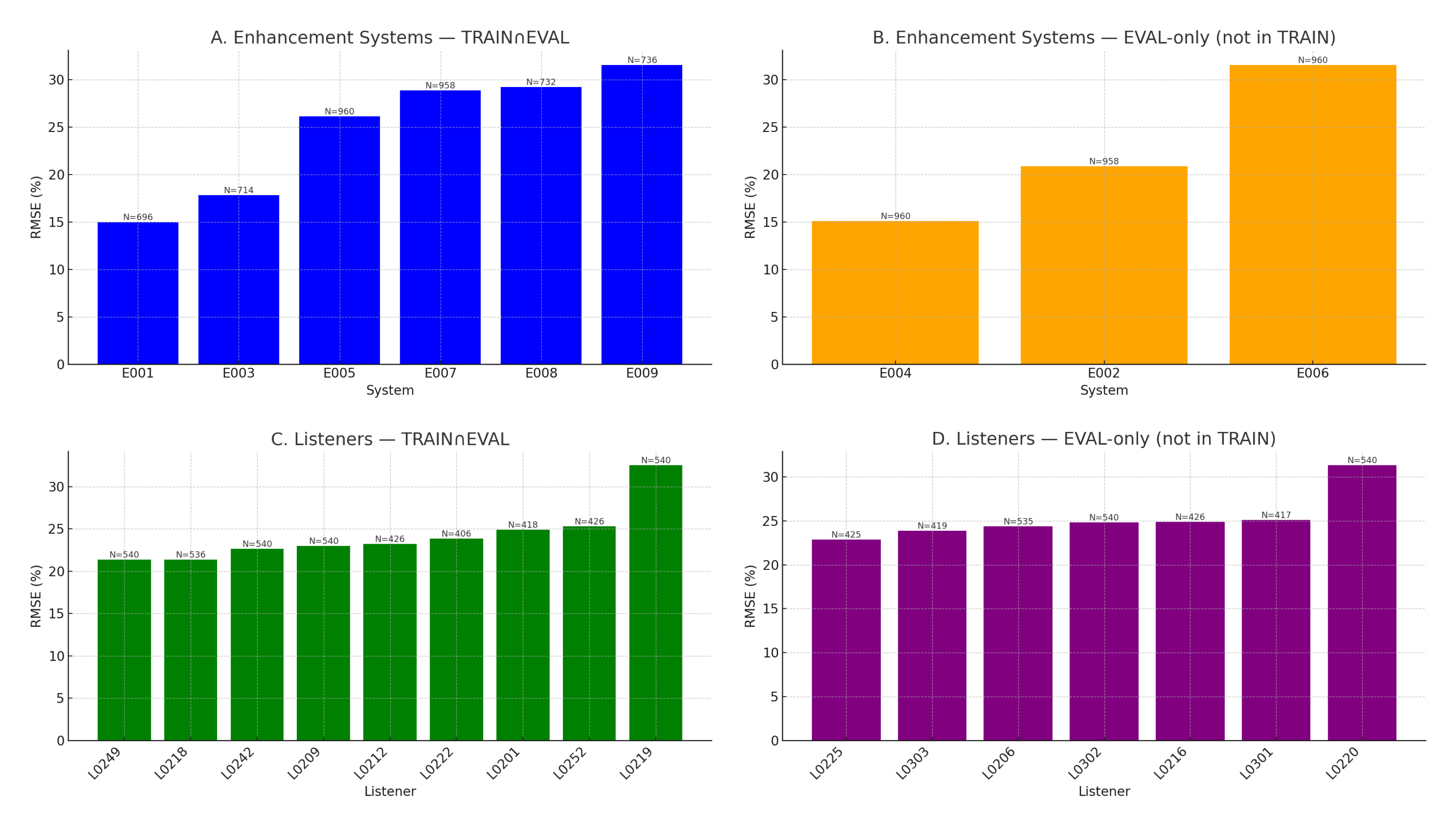}
  \caption{Error stratified by enhancement systems and listeners.
  (A) systems present in both Train and Eval; (B) Eval-only systems;
  (C) listeners seen in Train; (D) Eval-only listeners. Bars show RMSE
  per identity with sample counts (N).}
  \label{fig:strata}
  \vspace{-2mm}
\end{figure}

\noindent\textbf{Enhancement vs.\ listener generalization.}
Weighted RMSE over the four groups in Fig.~\ref{fig:strata} are:
(A) 25.8130 ($N{=}4796$), (B) 23.5139 ($N{=}2878$),
(C) 24.4725 ($N{=}4372$), (D) 25.6265 ($N{=}3302$).
Two observations emerge.
(i) Overlap with Train does not help; errors are higher on overlapping systems than on Eval-only systems. (ii) Listeners: predictions are better for listeners seen during training.

\begin{figure}[t]
  \centering
  \includegraphics[width=.80\columnwidth]{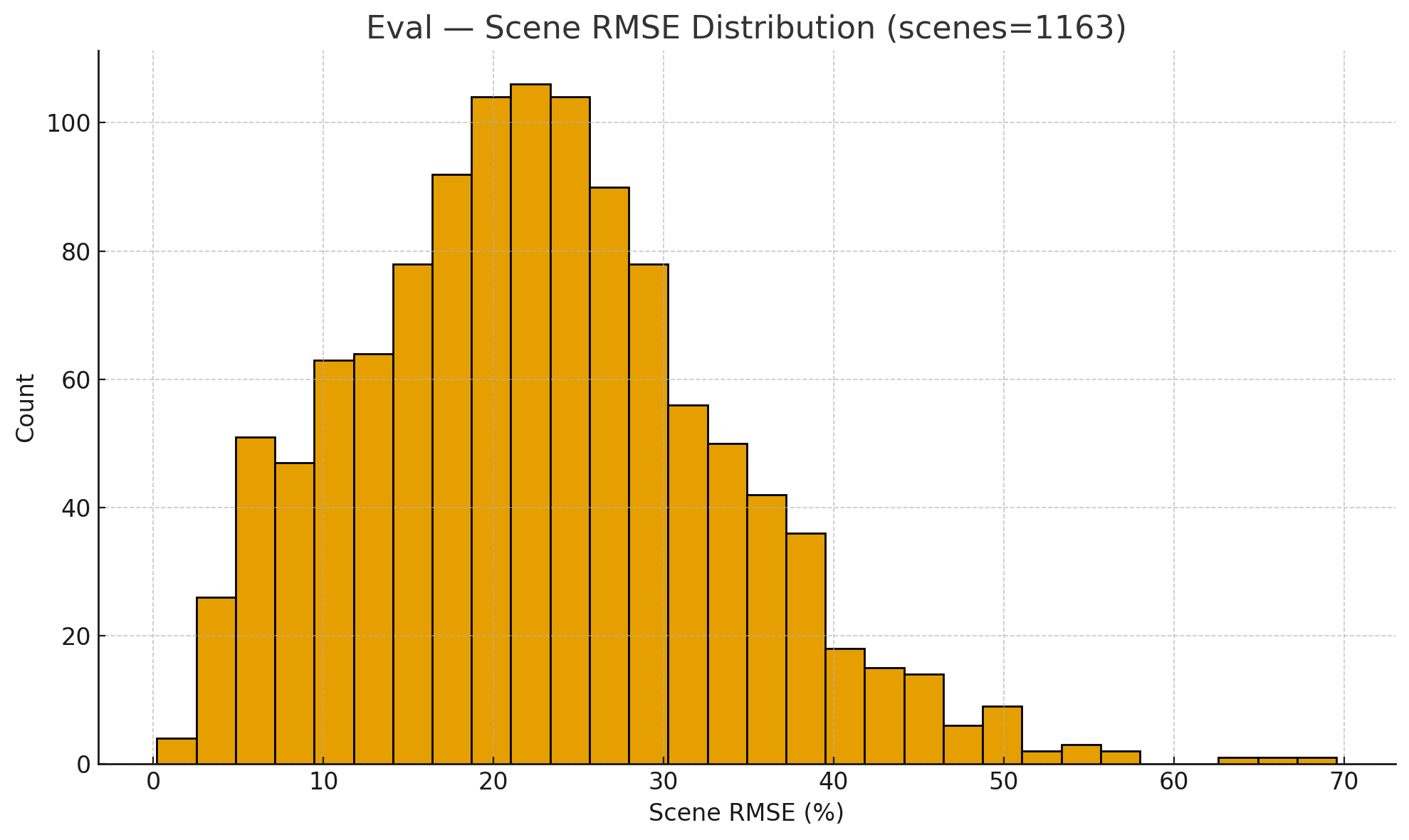}
  \caption{Eval scene-wise RMSE distribution ($N{=}1163$).}
  \label{fig:scene-hist}
  \vspace{-2mm}
\end{figure}

\noindent\textbf{Scene-wise error distribution.}
The Eval histogram in Fig.~\ref{fig:scene-hist} (1,163 scenes) is right-skewed: most scenes lie in 15–30\% RMSE, but a small tail ($\sim$6\%) exceeds 40\%, with a few $>$50\%. These hard scenes disproportionately raise the overall RMSE. Future work will profile the right tail across system, noise type, SNR, reverberation, and listener factors.

\section{Conclusion}
We introduced an intrusive SFM-based intelligibility predictor that builds on the CPC2-winning recipe by adding explicit reference conditioning, mid–deep layer selection (10–16) and an MSCNN front end. 
Systematic ablations showed how reference conditioning and cross-ear modeling help and aggregating mid–deep layers outperforms single-layer or mean/CLS pooling. 
Our system ranked \#1 on CPC3 (Dev RMSE 22.36; Eval RMSE 24.98). 
Although the runner-up reported a lower Dev RMSE, our method achieved better Eval RMSE, indicating stronger generalization and stability.

%\vfill
%\pagebreak

\bibliographystyle{IEEEbib}
\bibliography{strings,refs}

\end{document}